\documentclass[a4paper]{spie}  %>>> use this instead for A4 paper
\usepackage{graphicx}
\usepackage{xspace}
\usepackage[colorlinks=true,citecolor=blue]{hyperref}

\newcommand{\mic}{$\mu$m\xspace}

\newcommand{\lsd}{\hbox{$\lambda/D$}\xspace}

\title{Characterisation of a turbulent module for the MITHIC high-contrast imaging testbed}

\author{A. Vigan\supit{a}, M. Postnikova\supit{a}, A. Caillat\supit{a}, J.-F. Sauvage\supit{a,b}, K. Dohlen\supit{a},  K. El Hadi\supit{a}, \\
T. Fusco\supit{a,b}, M. Lamb\supit{c}, M. N'Diaye\supit{e}
\skiplinehalf
\supit{a} Aix Marseille Universit\'e, CNRS, LAM (Laboratoire d'Astrophysique de Marseille) UMR 7326, 13388, Marseille, France \\
\supit{b} ONERA, The French Aerospace Lab BP72, 29 avenue de la Division Leclerc, 92322 Ch\^atillon Cedex, France \\
\supit{c} University of Victoria, 3800 Finnerty Rd, Victoria, Canada \\
\supit{d} NRC Herzberg Astronomy, 5071 W. Saanich Rd, Victoria, Canada \\
\supit{e} Space Telescope Science Institute, 3700 San Martin Drive, Baltimore MD 21218, USA \\
}

\authorinfo{\noindent Send correspondence to A. Vigan: arthur.vigan@lam.fr}

\begin{document}
\maketitle

%%%%%%%%%%%%%%%%%%%%%%%%%%%%%%%%%%%%%%%%%%%%%%%%%%%%%%%%%%%%%
\begin{abstract}
Future high-contrast imagers on ground-based extremely large telescopes will have to deal with the segmentation of the primary mirrors. Residual phase errors coming from the phase steps at the edges of the segments will have to be minimized in order to reach the highest possible wavefront correction and thus the best contrast performance. To study these effects, we have developed the MITHIC high-contrast testbed, which is designed to test various strategies for wavefront sensing, including the Zernike sensor for Extremely accurate measurements of Low-level Differential Aberrations (ZELDA) and COronagraphic Focal-plane wave-Front Estimation for Exoplanet detection (COFFEE). We recently equipped the bench with a new atmospheric turbulence simulation module that offers both static phase patterns representing segmented primary mirrors and continuous phase strips representing atmospheric turbulence filtered by an AO or an XAO system. We present a characterisation of the module using different instruments and wavefront sensors, and the first coronagraphic measurements obtained on MITHIC.
\end{abstract}

\keywords{Turbulence simulation, Zernike wavefront sensor}
%%%%%%%%%%%%%%%%%%%%%%%%%%%%%%%%%%%%%%%%%%%%%%%%%%%%%%%%%%%%%

\section{Introduction}
\label{sec:introduction}

Direct imaging of exoplanets requires both high-angular resolution, to be able to resolve the planet from its star, and high-contrast, to be able to detect the much fainter signal of the planet\cite{Oppenheimer2009}. In this field, the new generation of high-contrast imagers such as VLT/SPHERE\cite{Beuzit2008}, Gemini/GPI\cite{Macintosh2008} or Palomar/P1640\cite{Hinkley2008} have seen their first light\cite{Macintosh2014,Vigan2016} and now regularly produce high-quality scientific results\cite{Oppenheimer2013,Macintosh2014,Boccaletti2015}. However, the next generation of instruments on extremely large telescopes (ELTs) will have to deal with additional caveats, the first one being the segmentation of the primary mirrors of these telescopes. While it is reasonable to assume that the ELTs will be coarsely cophased at the level of a few hundreds of nanometers RMS using dedicated optical cophasing techniques\cite{Chanan1998,Chanan1999,Chanan2000,Gonte2009,Gonte2011,Surdej2010,Vigan2011}, high-contrast imaging requires an improved measure and control of the cophasing errors, at the level of a few nanometers RMS.

To investigate these problems as well as develop new high-contrast imaging and wavefront sensing concepts, we have developed the Marseille Imaging Testbed for HIgh Contrast (MITHIC) at \emph{Laboratoire d'Astrophysique de Marseille} (LAM). MITHIC was developed\cite{NDiaye2012} over the past 5 years to validate wavefront control strategies such as the Zernike sensor for Extremely accurate measurements of Low-level Differential Aberrations (ZELDA)\cite{NDiaye2013} and COronagraphic Focal-plane wave-Front Estimation for Exoplanet detection (COFFEE)\cite{Paul2013}.

To continue these developments in the context of the new generation ELTs, we have recently acquired an atmospheric turbulence simulation module that offers both static phase patterns representing segmented primary mirrors and continuous phase strips representing atmospheric turbulence filtered by an AO or an XAO system. In this work, we present the characterization of the module using different instruments and wavefront sensors, and the first coronagraphic measurements obtained on MITHIC. In Sect.~\ref{sec:definition_phase_screen} we present the definition and manufacturing of the phase screen, in Sect.~\ref{sec:optical_characterization} the characterization using an interferometer and an interferometric microscope, and in Sec.~\ref{sec:first_results_mithic} the first results obtained on the MITHIC testbed with ZELDA and with a Roddier coronagraph.

\section{Definition and manufacturing of the phase screen}
\label{sec:definition_phase_screen}

\subsection{Definition}
\label{sec:definition}

\begin{figure}
  \centering
  \includegraphics[width=1\textwidth]{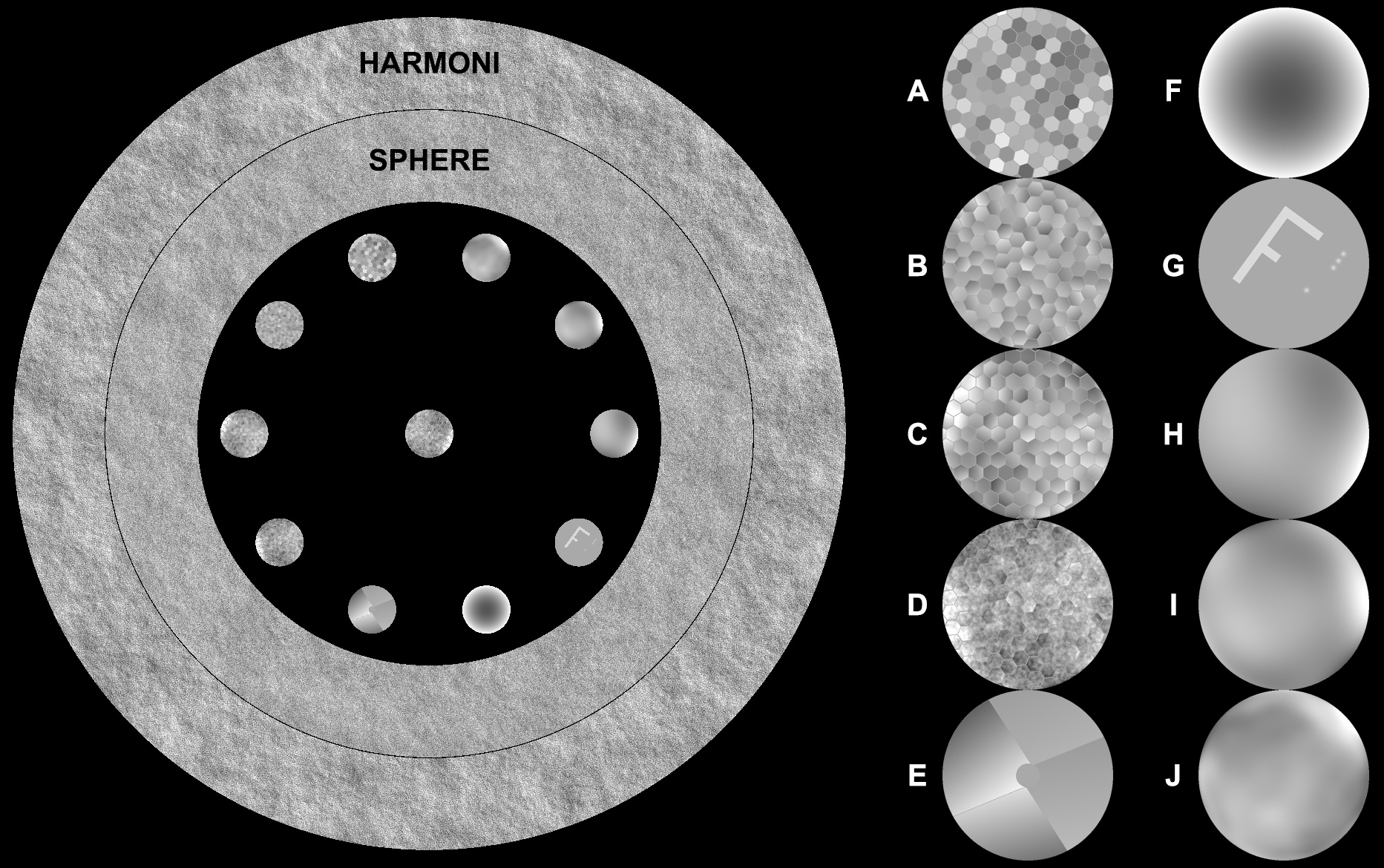}
  \caption{Complete definition of the phase screen. The outer part contains two strips with simulated AO-filtered turbulence residual phase corresponding to VLT/SPHERE-like and E-ELT/HARMONI-like systems. The inner part contains 11 static phase patterns corresponding to the following cases: (A) segments with piston; (B) segments with piston, tip and tilt; (C) segments with piston, tip, tilt and low-order NCPA; (D) segments with piston, tip, tilt, low-order NCPA and AO-filtered turbulance for a E-ELT/HARMONI-like system; (E) low-wind effect on a VLT-like pupil; (F) pure focus; (G) influence functions; (H) NCPA 10 terms; (I) NCPA 36 terms; (J) NCPA 200 terms. See text for more details.}
  \label{fig:definition}
\end{figure}

The phase screen was defined to serve different purpose in the high-contrast R\&D activities on-going at Laboratoire d'Astrophysique de Marseille (LAM): investigation of the VLT/SPHERE ultimate performance, development of high-contrast instrumentation for E-ELT instruments, and in particular for the HARMONI projects, development of focal-plane wavefront sensing and co-phasing techniques. In order to be more versatile, we included simulation of both AO-filtered residual turbulence and various static phase patterns.

For the residual turbulence, we considered two cases representing an extreme AO system (similar to VLT/SPHERE) and a more conventional AO system (similar to E-ELT/HARMONI). For the static patterns, the following cases were considered:

\begin{itemize}
    \item \textbf{Non-common path aberrations (NCPA)}: Zernike coefficients randomly drawn according to a decreasing profile in 1/radial\_order$^2$. This behaviour is representative of typical NCPA residuals. Three different patterns are included in the phase screens, with 10, 36 and 200 Zernike modes;
    \item \textbf{segments}: simulated 10$\times$10 segemented telescope with 4\% gap. At the scale of the VLT, this simulates 0.8~m segments with 3.2~cm gap, while at the scale of the E-ELT, this simulates 3.7~m segments with 15~cm gap. While not realistic in the latter case, this is the best that could be done with a 5~mm pupil on MITHIC and 10~\mic pixel size (see Sec.~\ref{sec:manufacturing}). Four different patterns were included with different contributions: piston, tip, tilt, low-order NCPA and AO-filtered turbulence with an E-ELT/HARMONI like system;
    \item \textbf{low-wind effect (LWE)}: the low-wind effect is a dome-seeing effect that affects the VLT when very low external wind speeds are on-going\cite{Sauvage2016}. The main effect is to create differential piston between parts of the VLT pupil separated by the spider vanes (see Sauvage et al., this conference). The pattern included in the phase screen correspond to typical values that have been observed in real observing conditions with the VLT/SPHERE instrument;
    \item \textbf{influence function}: a specific pattern was included with influence functions for a 70$\times$70 actuator deformable mirror. This pattern also includes an "F" to determine the orientation of the spatial axes in the system.
\end{itemize}

\noindent The complete definition of the phase screens, including detailed view of the static patterns, is given in Fig.~\ref{fig:definition}.

The amplitude and spatial scale of the filtered turbulence and static patterns were calculated to correspond to the desired physical properties, while taking into account the size and properties of the phase screens as well as the setup of the MITHIC bench:

\begin{itemize}
    \item pupil diameter: 5~mm
    \item wavelength: 677~nm
    \item phase screen diameter: 100 mm
    \item phase screen pixel size: 10~\mic
\end{itemize}

\subsection{Manufacturing}
\label{sec:manufacturing}

\begin{figure}
  \centering
  \includegraphics[width=0.7\textwidth]{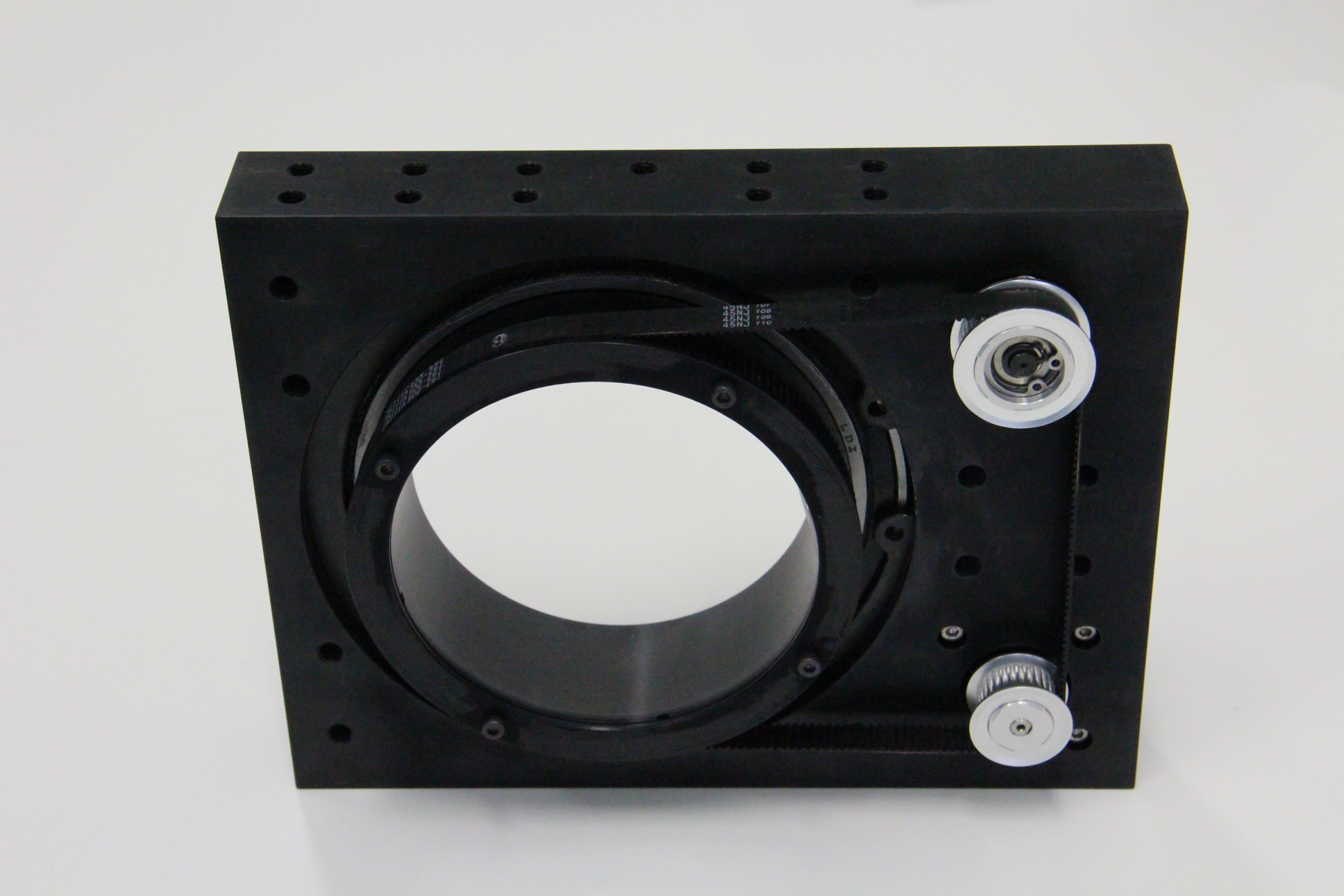}
  \caption{Phase screen in its rotating opto-mechanical mount as delivered by SILIOS. The screen is designed to work in transmission. Rotation is performed using a stepper motor controlled over USB using a dedicated software.}
  \label{fig:mount}
\end{figure}

The phase screen and it rotating mechanical mount were provided by the SILIOS Technologies SA, a company created in 2001 in Peynier (near Aix-en-Provence, France). The company is expert in passive micro-structured optics. SILIOS takes benefit from semiconductor manufacturing equipments, processes and methods to produce its optical components in clean room (ISO 5 to 3). The main part of its production is diffractive optics manufactured using a cumulative etching technology.

The cumulative etching technology allows reaching multilevel stair-like topologies directly into fused silica substrates\cite{Caillat2014}. Such surface topologies are used to produce phase functions for gratings, diffractive laser beam shapers and turbulence phase screens for example. The technology is based on successive masking photolithography and reactive ion etching steps (i.e. successive etching steps through resin masks). Thus, the whole phase pattern is engraved at the same time, resulting in a very high uniformity over the full engraved area.

The cumulative etching of N mask levels allows reaching depth profile discretized over $2^N$ steps. For instance, a process involving six levels (i.e. four masking photolithography and etching individual steps) leads to a 64 phase level profile. The etched depth of each level is given by:

\begin{equation}
    d_{i} = d_{max}\frac{2^{N-1}}{2^{N}-1}\frac{1}{2^{N-i}}
\end{equation}

where $i$ is the number of the mask level, $d_{max}$ is the maximum achievable depth of the profile, and $N$ is the total number of mask levels.

The photolithography tool is a contact photo-masker with an UV400 source. The reachable critical dimension (smallest lateral feature size) is about 1~\mic, but for our phase screen we used 10~\mic pixel size, which is sufficient for our laboratory setup. To provide a smooth variation of the phase, six masks are used, providing 64 discretization levels.

Two main manufacturing inaccuracies can lead to distortion of the profile and thus to the reduction of its efficiency: inaccuracy on the etched depths and inaccuracy on the alignment of each mask pattern with the others:

\begin{itemize}
    \item The etching accuracy on each depth level has to be much lower than the smallest step to avoid strong distortion in the profile. Specific fused silica processes have been developed by SILIOS during the past 10 years with an etching accuracy in the range of $\pm$5 to 10~nm insuring a very small profile deviation. To control the depths, several test patterns are positioned all around the grating useful aperture on each mask. These test patterns give access to the measurement of each individual etching depth, as well as some cumulative etching depths. The depth for one mask is updated from the previous mask depth measured. Therefore these test patterns allow controlling every etching depth during the process and at the end of the manufacturing task. The depth measurements are performed using a contact mechanical profilometer, whose accuracy is $\pm$3~nm.
    \item The inaccuracy on the alignment of each mask pattern with the others is potentially a much bigger source of error. Misalignment in between masks levels produces unwanted hollows or spikes located on the edges of the steps. This issue was previously identified in the manufacturing of high-precision gratings for the Euclid mission\cite{Caillat2014}. For our phase screens, these sharp features will create high-spatial frequency artifacts in the coronagraphic images.
\end{itemize}

The phase screen was provided with an opto-mechanical mount designed by SILIOS, which includes a rotating part and a stepper motor that allows a steps below $10^{-4}$ turn/step. Continuous rotation is possible between 0.1 and 6 turn/min. The motor is controlled over USB using a dedicated software provided by SILIOS.

\section{Interferometric characterization of the phase screens}
\label{sec:optical_characterization}

\subsection{Interferometer}
\label{sec:interferometer}

\begin{figure}
  \centering
  \includegraphics[width=0.6\textwidth]{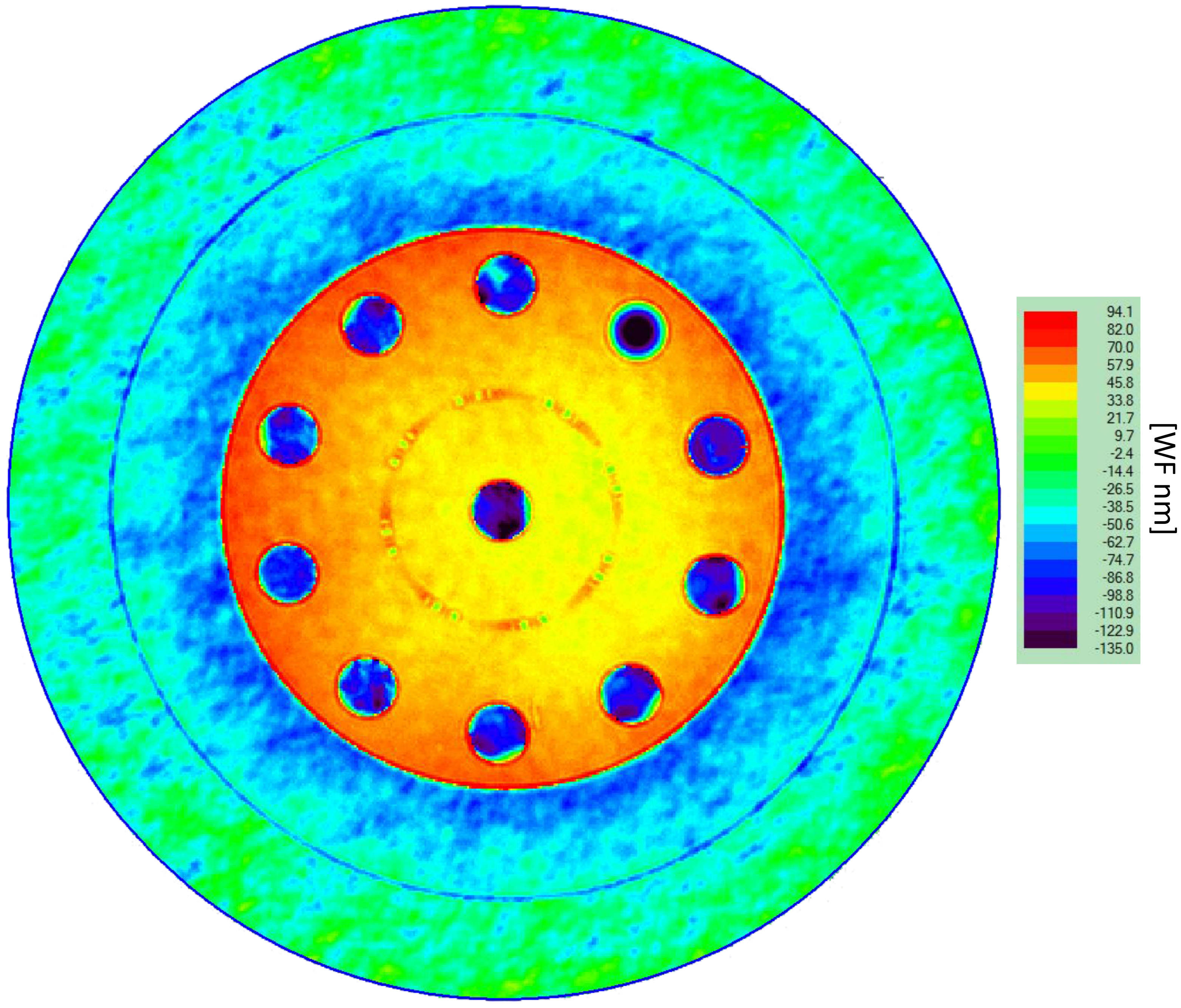}
  \caption{Interferogram covering the full phase screen. The spatial resolution is approximately 200~\mic/pixel.}
  \label{fig:interferogram}
\end{figure}

Before being engraved with the phase patterns, the fused silica plate was characterized by SILIOS with a Zygo interferometer. They measured an optical surface quality of $\lambda/37$~rms, and a parallelism of 0.3''.

After the phase screen was received at LAM, a first measurement was performed with a M\"oller interferometer (Fig.~\ref{fig:interferogram}). With a spatial resolution of $\sim$200~\mic/pixel, this type of interferogram does not provide a spatial resolution sufficient to measure accurately the static patterns or the simulated turbulence, but it provides a good view of the general optical quality of the phase screen. We can clearly identify the two AO-filtered turbulence regions on the outer part of the phase screen, as well as the 11 static phase patterns. The circular markings around the center correspond to markings used by SILIOS to align the manufacturing masks and to control the depth of the etching at each step of the manufacturing process. Measurements of the flatness in the central, non-engraved area (yellow part on Fig.~\ref{fig:interferogram}) are compatible with the values reported by SILIOS before etching.

\subsection{Interferometric microscope}
\label{sec:interferometric_microscope}

\begin{figure}
  \centering
  \includegraphics[width=0.8\textwidth]{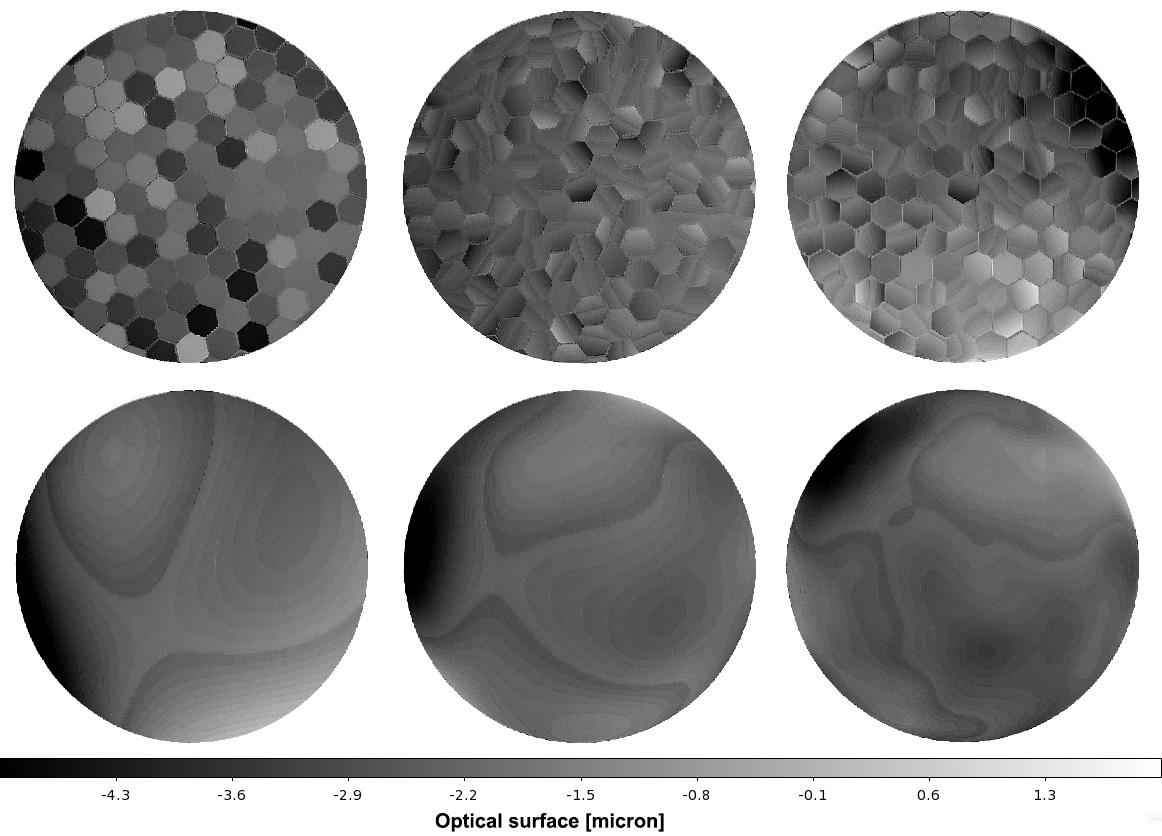}
  \caption{Measurements of static patterns using an interferometric microscope. Top, from left to right: segments with piston; segments with piston, tip and tilt; segments with piston, tip, tilt and low-order NCPA. Bottom, from left to right: NCPA with with 10, 36 and 200 Zernike modes. The contour lines corresponding to the levels engraved by the manufacturing process are clearly visible in these measurements. The spatial resolution of these data is $\sim$3.65~\mic/pixel.}
  \label{fig:screens_microscope}
\end{figure}

The phase screen was then measured using an interferometric microscope. We acquired measurements of all of the static patterns and the turbulence with the Vertical Shifting Interfometer (VSI) mode in white light, which provides a spatial resolution of 3.65~\mic and a vertical accuracy of $\sim$20~nm. The measurements were done over 5$\times$5~mm$^2$ areas (using stitching) to cover the full patterns. In addition, we also acquired high-vertical accuracy ($<1$~nm) and high-spatial resolution (0.2~\mic) measurements with the Phase Shifting Interferometer mode (PSI, green light) on smaller areas ($\sim$100~$\mu \mathrm{m}^2$).

\begin{figure}
  \centering
  \includegraphics[width=0.8\textwidth]{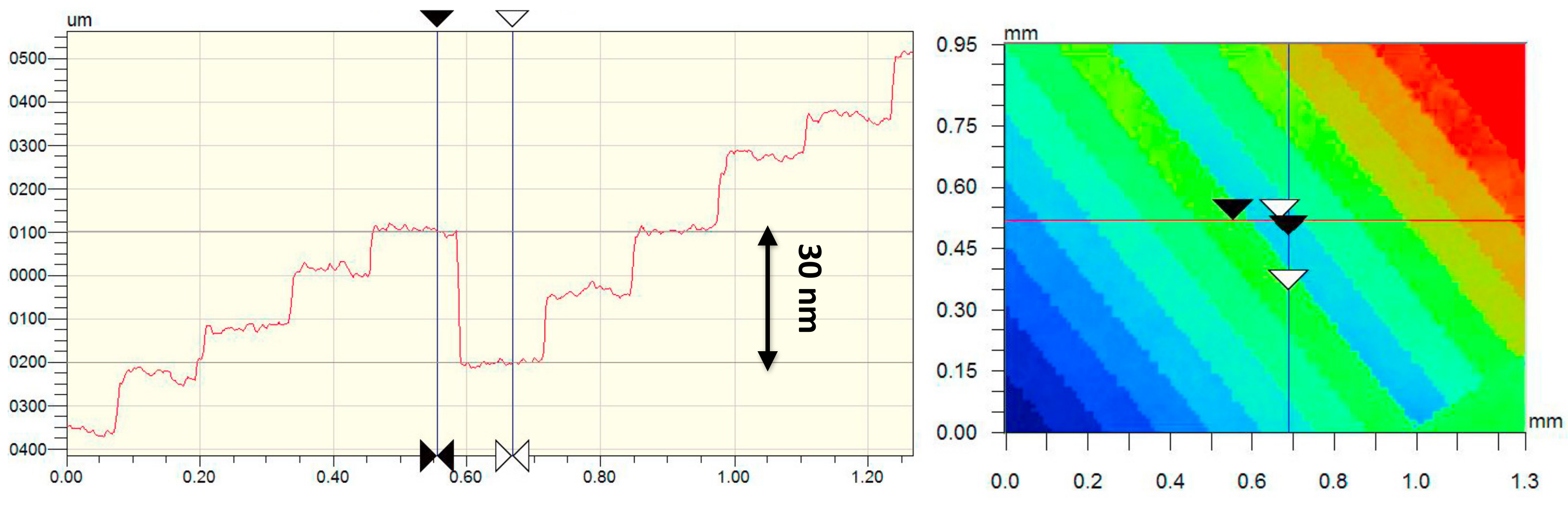}
  \caption{Zoom on the 30~nm manufacturing defect that appears in areas with continuous specified slopes. The origin of the problem is under investigation by SILIOS.}
  \label{fig:defect}
\end{figure}

Global measurements of 6 static patterns in VSI mode are presented in Fig.~\ref{fig:screens_microscope}. In these measurements, the pixelated structure inherent to the manufacturing process using binary masks is clearly visible, in particular at the edges of the segments. The contour lines corresponding to the different etching levels are also well defined and visible in the NCPA patterns. Within these images, we detected a possible manufacturing defect that appears in areas with continuous specified slopes, which would correspond to one etching level not deep enough. At the level of the surface, it translates into a jump of $\sim$30~nm (Fig.~\ref{fig:defect}). The problem is being investigated by SILIOS to understand its origin and how to solve it. Since it probably corresponds to one level that has not been etched deep enough, a possible solution would be to reprocess the phase screen with the corresponding manufacturing mask, etching an additional 30~nm.

\begin{figure}
  \centering
  \includegraphics[width=0.6\textwidth]{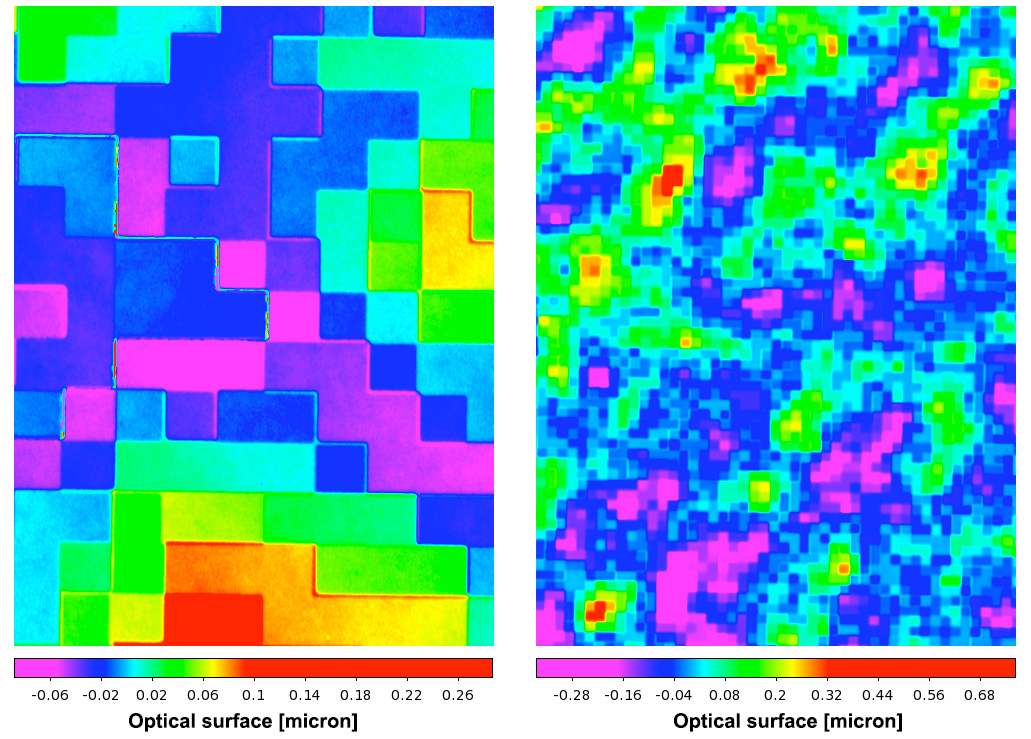}
  \caption{Detailed measurements in PSI mode inside the AO-filtered turbulence residuals corresponding to a VLT/SPHERE-like instrument. The spatial resolution in the right and left images is respectively 0.2 and 1.0~\mic/pixel.}
  \label{fig:screens_microscope_detail}
\end{figure}

\begin{figure}
  \centering
  \includegraphics[width=0.8\textwidth]{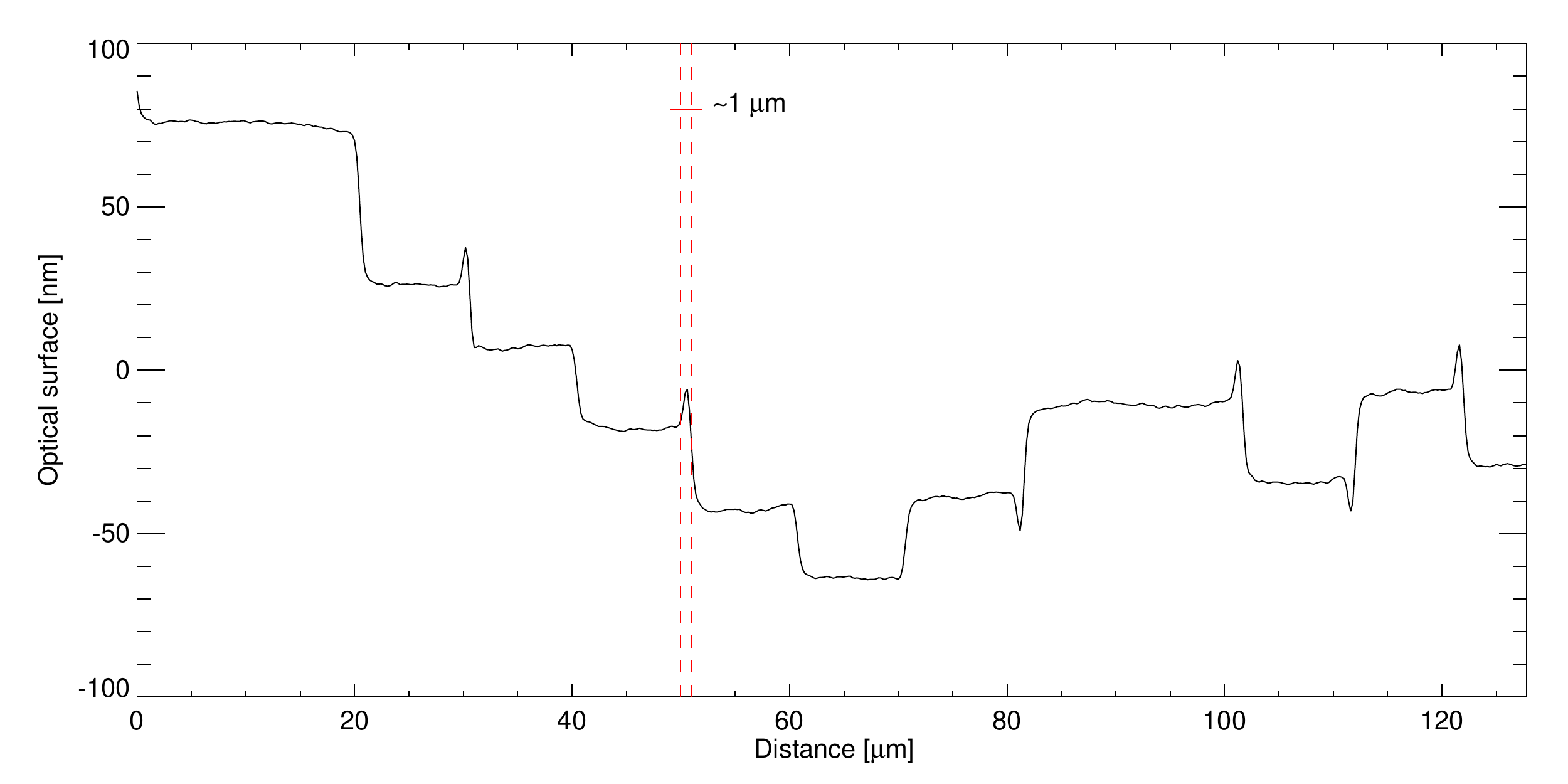}
  \caption{Vertical cut in the detailed measurement presented in the left of Fig.~\ref{fig:screens_microscope_detail}. This cut shows one of the typical manufacturing inaccuracy, which creates spikes and hollows at the edges of pixels because of small errors on the manufacturing masks alignment. The size of such spikes/hollows is $\sim$1~\mic.}
  \label{fig:cut}
\end{figure}

A much finer view taken inside the AO-filtered turbulence residuals corresponding to a VLT/SPHERE-like instrument is visible in Fig.~\ref{fig:screens_microscope_detail}. In these data, the pixel structure is clearly apparent and we can see one of the typical manufacturing inaccuracies presented in Sect.~\ref{sec:manufacturing}: a slight misalignment of the manufacturing masks will create spikes and hollows at the edge of the pixels that define the phase profile. This is illustrated in Fig.~\ref{fig:cut} where we present a vertical cut of the fine measurement presented in Fig.~\ref{fig:screens_microscope_detail} (left). The spikes/hollows have a size of $\sim$1~\mic, corresponding to an equivalent misalignment of the mask during manufacturing. Such defects were anticipated and should not have a significant impact for our high-contrast applications since they correspond to very high-spatial frequencies that will create structures in the focal plane images very far from the optical axis.

\section{First results on MITHIC}
\label{sec:first_results_mithic}

\subsection{Presentation of the bench}
\label{sec:presentation_bench}

\begin{figure}
  \centering
  \includegraphics[width=0.7\textwidth]{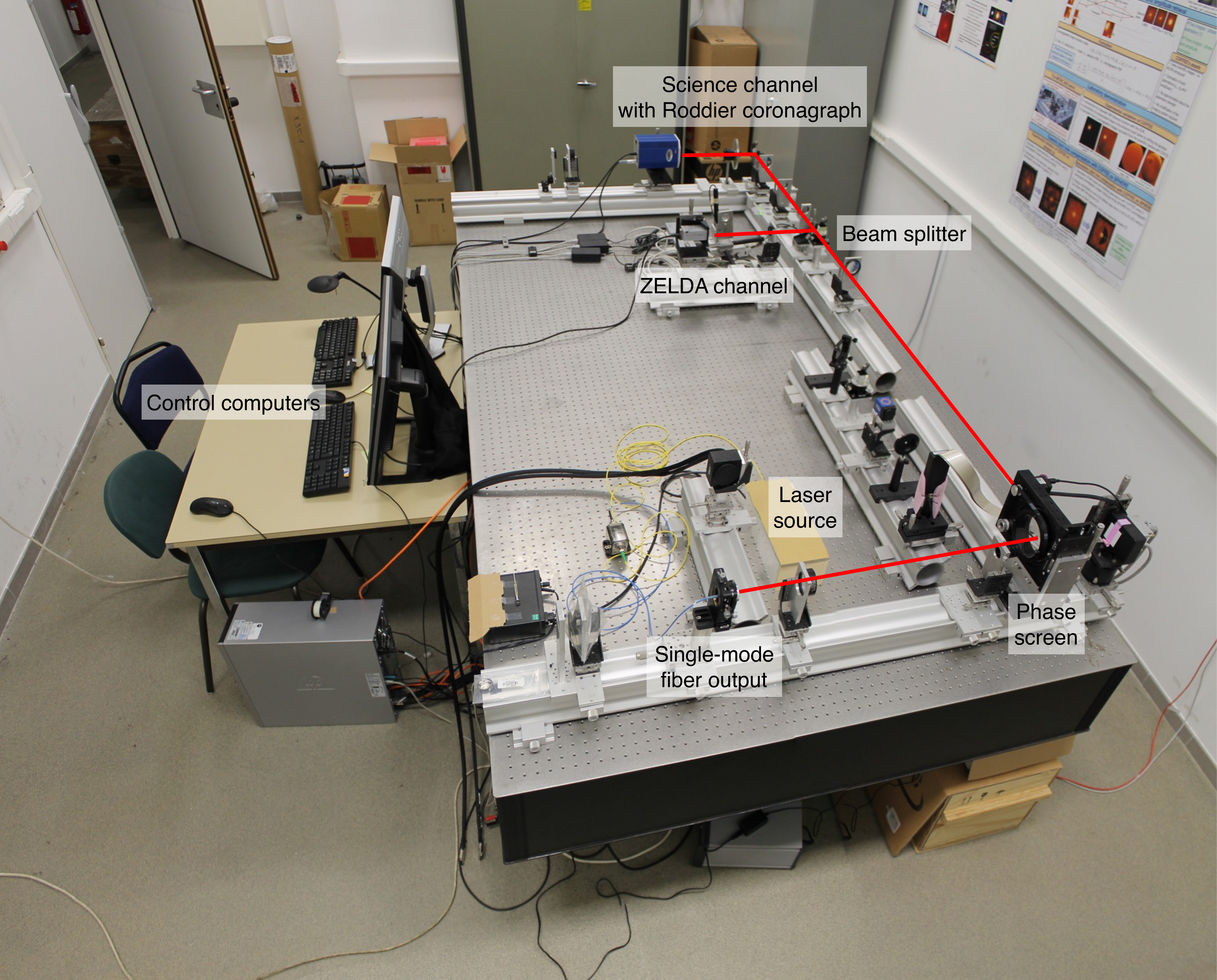}
  \caption{Setup of the MITHIC bench at LAM.}
  \label{fig:mithic_setup}
\end{figure}

The MITHIC testbed was previously described in N'Diaye et al. (2012)\cite{NDiaye2012}, but we briefly describe the setup that was used for the characterization of our phase screens. The light source on the bench is a laser diode emitting at 677~nm through a single mode fiber. The end of the fiber is placed in the focal plane of high-quality lens, which produces a collimated beam into which we insert a diaphragm of diameter 5~mm, which defines the pupil of our system, with respect to which the rest of the bench is aligned. The 5~mm collimated beam is then sent to another lens after which we insert a beam-splitter. The splitter splits the beam between the ZELDA channel and the coronagraphic channel. In the ZELDA channel, we have a Zernike phase mask of radius $\sim$\lsd in the focal plane of the beam. The mask is designed to convert the small phase aberrations into intensity variations (see N'Diaye et al. 2013\cite{NDiaye2013} and Sec.~\ref{sec:zelda_measurements}) by introducing a phase shift of $\pi/2$. After the mask, another lens allows reimaging the pupil on a PixeLINK camera. In the coronagraphic channel, we insert a Roddier coronagraphic phase mask in the focal plane. Then a lens allows to produce again a collimated beam in which a Lyot stop is placed in the plane conjugated with the entrance pupil. A final lens is then used to produce a focal plane image of the PSF on a Photometrics CoolSNAP HQ$^2$ camera. The full setup is illustrated in Fig.~\ref{fig:mithic_setup}.

\subsection{ZELDA measurements}
\label{sec:zelda_measurements}

The ZELDA concept\cite{NDiaye2013} has been originally proposed to solve the problem of non-common path aberations (NCPA) at the level of the coronagraph, since these aberrations have been identified as a significant limitation for the detection of planets through direct imaging at very small angular separations. N'Diaye et al. (2013)\cite{NDiaye2013} proposed the use of a Zernike phase mask sensor to calibrate the NCPA seen by the coronagraph in exoplanet direct imagers. This phase-contrast method uses a phase-shift mask to modulate the phase differential aberrations into intensity variations in the pupil plane. Since differential aberrations in exoplanet imagers are small, a linear or quadratic relation between the wavefront errors and the pupil intensity enables reconstructing the differential aberrations at nanometric accuracy with a simple, fast algorithm, making calibration in real time possible. 

Zernike sensors have been explored in astronomy to address various instrumentation aspects, such as wavefront sensing in adaptive optics systems or cophasing of telescope segmented primary mirror\cite{Bloemhof2003,Bloemhof2004a,Dohlen2004,Surdej2010,Wallace2011,Vigan2011}. Recently, the Zernike sensor has been adopted for the WFIRST mission to measure low-order aberrations in its coronagraphic instrument and control pointing errors and focus drifts on the coronagraphic mask\cite{Spergel2013a,Spergel2015,Zhao2014}. Laboratory demonstration of the concept have been carried out in this context\cite{Shi2015}. We have also performed preliminary tests and obtained encouraging results of the Zernike sensor on the coronagraphic testbed in Marseille\cite{NDiaye2012,NDiaye2014b,Dohlen2013}. Finally, we recently achieved an important milestone by validating the ZELDA concept for the measurement and correction of NCPA inside VLT/SPHERE\cite{NDiaye2016}.

\begin{figure}
  \centering
  \includegraphics[width=0.7\textwidth]{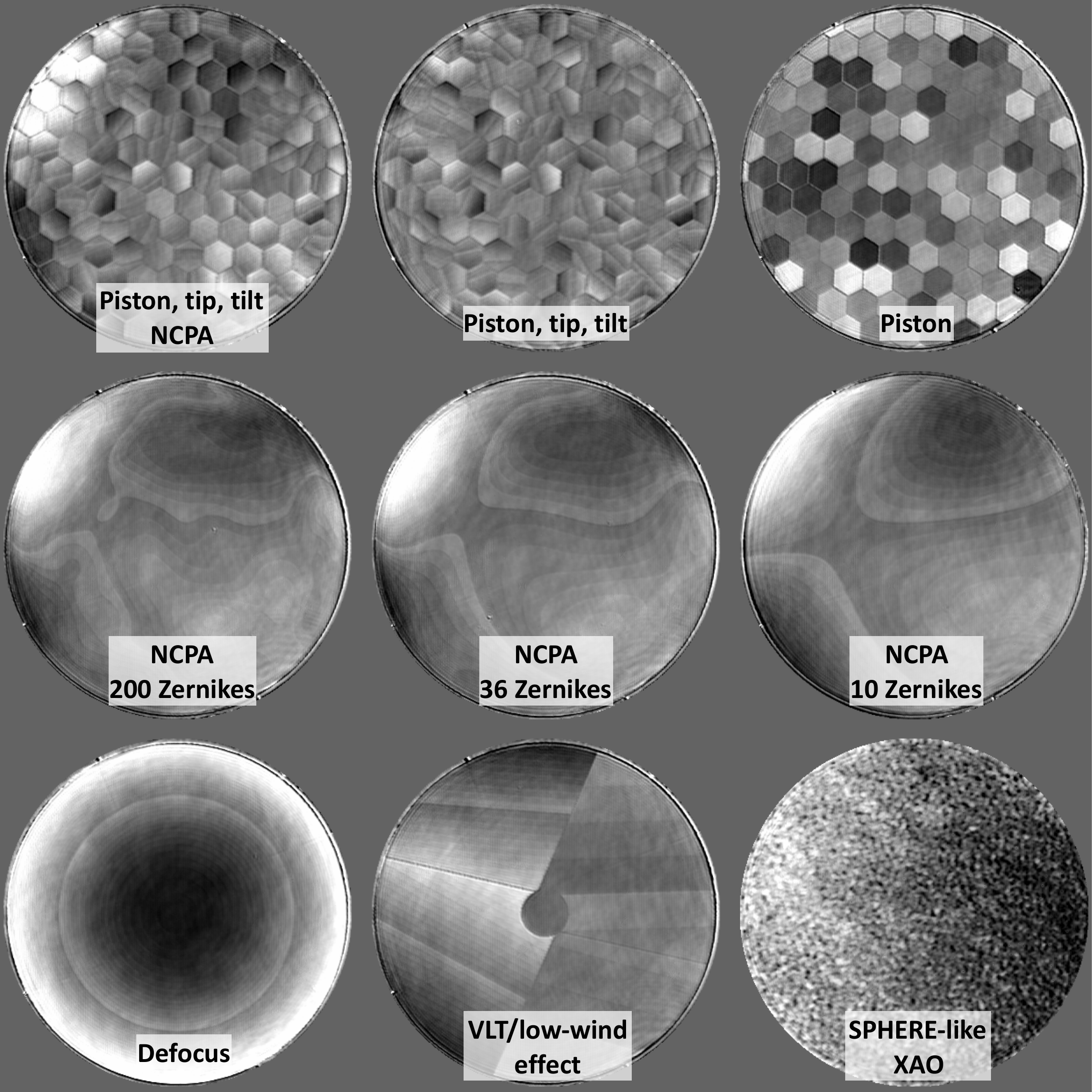}
  \caption{ZELDA measurements of some of the static patterns and one area in the VLT/SPHERE-like filtered turbulence.}
  \label{fig:zelda}
\end{figure}

\begin{figure}
  \centering
  \includegraphics[width=0.5\textwidth]{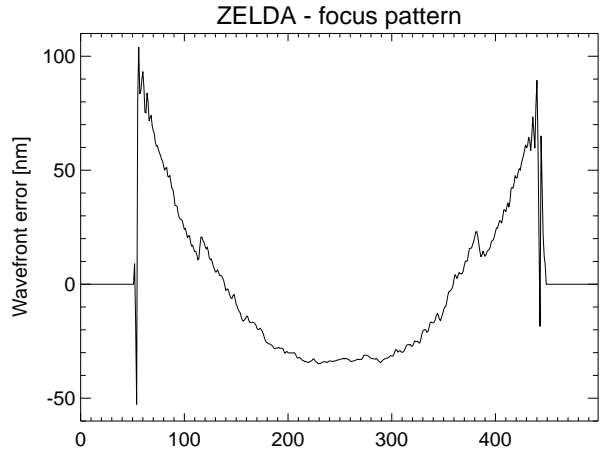}
  \caption{Cut through the OPD map measured with ZELDA on the defocus pattern. The manufacturing defect identified in Sec.~\ref{sec:interferometric_microscope} is visible as a $\sim$15~nm jump on the wavefront on both sides of the focus pattern. It is also easily identifiable in Fig.~\ref{fig:zelda} as an annulus in the focus pattern.}
  \label{fig:zelda_focus}
\end{figure}

The ZELDA sensor on MITHIC was used to obtain data with our phase screen. We refer the reader to N'Diaye et al. (2013, 2016)\cite{NDiaye2013,NDiaye2016} for the full theoretical treatment and the data analysis. For the MITHIC measurement, we acquired data on the static patterns and turbulence areas, as well as on a flat area of the phase screen. This latter measurement is used as a reference to measure the static aberrations of the testbed. This reference is subtracted to the OPD maps obtained on the other areas to remove the contribution of the static patterns.

These high-spatial resolution measurements (190 pixels over the pupil) confirm the high-quality of the phase screen (Fig.~\ref{fig:zelda}). They also confirm the presence of the small defect identified in Sec.~\ref{fig:zelda}, which will hopefully be solved by etching again the corresponding manufacturing mask. Figure~\ref{fig:zelda_focus} presents a cut through the defocus pattern, where the defect is visible as a $\sim$15~nm jump on the wavefront.

\subsection{Focal-plane measurements}
\label{sec:focal_plane_measurements}

\begin{figure}
  \centering
  \includegraphics[width=1.0\textwidth]{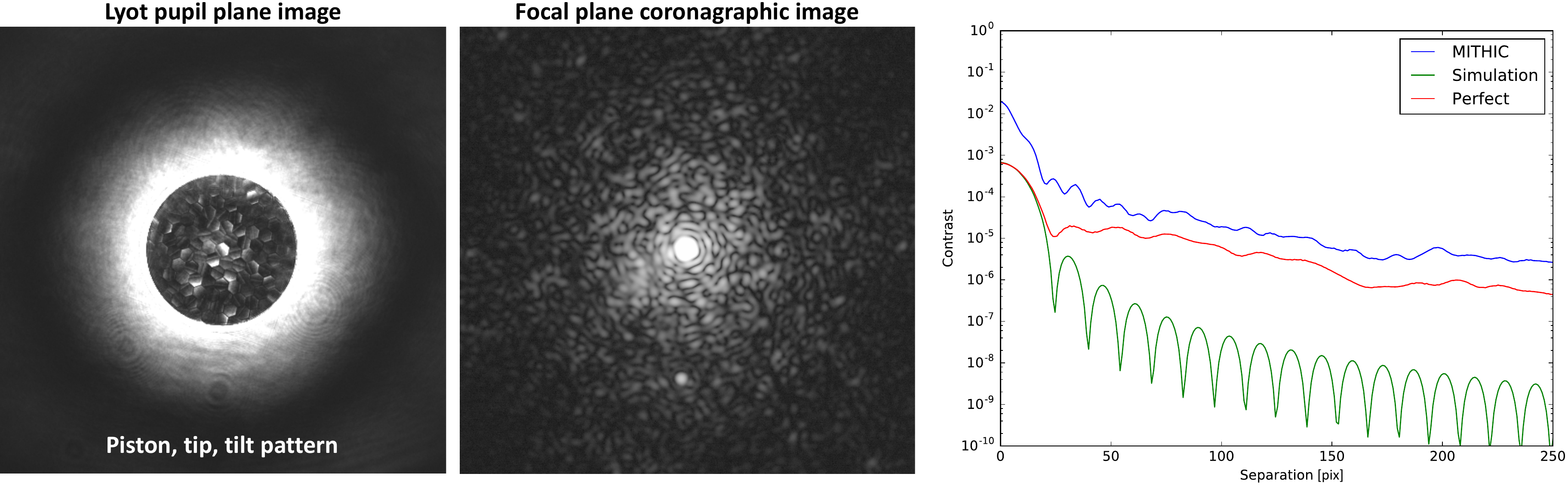}
  \caption{\emph{Left:} pupil-plane image of the piston-tip-tilt pattern in the plane of the Lyot stop, after a Roddier coronagraph. \emph{Center:} focal-plane coronagraphic image. \emph{Right:} Coronagraphic profiles corresponding to the measurement (blue), the theoretical profile with the specified phase pattern (red) and the theoretical performance of a perfect Roddier coronagraph (green).}
  \label{fig:coro_images}
\end{figure}

The MITHIC bench includes an apodized Roddier coronagraph, which we used in it non-apodized version to obtain first coronagraphic images with some of the static patterns. The results for the piston-tip-tilt pattern are presented in Fig.~\ref{fig:coro_images} where we see the Lyot plane pupil image downstream of the Roddier phase mask, the focal-plane coronagraphic image and a comparison of the coronagraphic profile with simulation. These initial measurements are encouraging for the future since we see that the measured profile is a factor $<$10 above the theoretical level obtained in simulation with the specified phase pattern. The main difference probably originates in the static aberrations of the testbed, which are not taken into account in the simulation.

\section{Prospects}
\label{sec:prospects}

The phase screen that we acquired from SILIOS Technologies are of very high-quality, despite a small defect that is currently under investigation and will hopefully be solved in the coming months. This phase screen is going to be used in several of the high-contrast imaging projects that are we are currently developing at LAM. These projects include:

\begin{itemize}
    \item dark hole generation with ZELDA
    \item coronagraphic performance in presence of residual segmentation errors
    \item coronagraphic phase diversity (COFFEE\cite{Paul2013}) estimation of segmentation errors
    \item combining ZELDA and COFFEE
    \item E-ELT/HARMONI: high-contrast applications with the pyramid WFS at LAM
\end{itemize}

%%%%%%%%%%%%%%%%%%%%%%%%%%%%%%%%%%%%%%%%%%%%%%%%%%%%%%%%%%%%%
\acknowledgments

AV acknowledges support from R\'egion Provence-Alpes-C\^ote d'Azur, under grant agreement 2014-02976, for the ASOREX project.

%%%%%%%%%%%%%%%%%%%%%%%%%%%%%%%%%%%%%%%%%%%%%%%%%%%%%%%%%%%%%
%%%%% References %%%%%
\bibliography{paper}
\bibliographystyle{spiebib}

\end{document}